\documentclass{article}

\usepackage{arxiv}

\usepackage[utf8]{inputenc}
\usepackage[T1]{fontenc}
\usepackage{amsmath, amssymb, amsfonts}
\usepackage{microtype}
\usepackage{hyperref}
\usepackage{natbib}

\title{Why AI Governance Frameworks Are Hard to Adopt: A Role-Based Stress Test of the NIST AI RMF}

\author{
  Joseph R. Simons \\
  Department of Engineering Management \& Systems Engineering \\ 
  The George Washington University, Washington, DC, USA\\
  \texttt{joseph.simons@gwmail.gwu.edu}
  \And
  David A. Broniatowski \\
  Department of Engineering Management \& Systems Engineering \\ 
  The George Washington University, Washington, DC, USA\\
  \texttt{broniatowski@gwu.edu}
}

\date{July 2, 2026}

\renewcommand{\shorttitle}{Why AI Governance Frameworks Are Hard to Adopt}

\begin{document}

\maketitle

\begin{abstract}
AI governance frameworks can be known, used, and implemented in form without becoming governance in practice. This paper examines that problem through a role-based stress test of the NIST Artificial Intelligence Risk Management Framework (AI RMF) in consumer lending. We treat framework adoption as a governance translation problem: whether RMF language can become role-usable, cross-level, authority-connected governance over the AI system-in-use, rather than producing governance-looking artifacts.

The study uses LLM-based role simulation as a structured analytic probe. We apply a 4 $\times$ 2 $\times$ 3 design across four organizational roles, two AI deployments, and three governance hard cases, producing 120 scored responses. Results show that local translation was not the main problem. Simulated actors generally understood their assigned roles and translated the RMF into local activity. The harder problem was whether that activity became governance value. Actor role was strongly associated with Cross-Level Governance Value, Authority Connection, Governance Translatability, and governance value. Deployment was strongly associated with Structural Fit: the RMF fit a bounded ML underwriting model more cleanly than a workflow-embedded LLM underwriting copilot. Risk reduction was harder still. It appeared only when governance value was present and Structural Fit was full, but neither condition was sufficient by itself.

The paper contributes a diagnostic account of framework-based AI governance. Frameworks create value when they help organizations see, interpret, escalate, authorize, and correct risk in the AI system-in-use. They also create value when they reveal limits of governability under existing evidence paths, authority structures, and system boundaries.
\end{abstract}

\section{Introduction}
\label{sec:introduction}

AI governance has an adoption puzzle. Governance frameworks are meant to translate broad principles into practical organizational action. Several AI governance frameworks are now well known, yet awareness of a framework does not mean that the framework has been adopted in any meaningful governance sense. Organizations may know the framework, map activities to it, and produce framework-shaped artifacts while still failing to change what they can see, decide, escalate, correct, or prevent.

This puzzle is visible in the NIST AI Risk Management Framework. The AI RMF is practical, widely recognized, and explicitly sociotechnical. It gives organizations a shared language for governing, mapping, measuring, and managing AI risk \citep{nist2023airmf}. It has also been promoted in practitioner guidance for organizations seeking to operationalize AI risk management \citep{deloitte2023nist}. Yet prior work suggests a gap between awareness and adoption \citep{kolo2025framework}. If the problem were only awareness, training, or documentation, that gap should be easier to close. The harder possibility is that framework adoption depends on governance conditions the framework does not create by itself.

This paper treats the adoption gap as a governance translation problem. Frameworks do not govern AI systems. Organizations do. A framework creates value only when its language travels through roles, responsibilities, evidence paths, authority structures, and correction mechanisms. Local implementation is therefore not enough. The question is whether framework-guided activity becomes governance inside the organization that must act on it.

The problem is also a system-boundary problem. AI risk is not always located in the model or tool artifact alone. In consequential organizational settings, risk may arise from the AI system-in-use: the model or tool as embedded in workflow, human reliance, documentation routines, supervision, incentives, escalation channels, and authority structures. A framework may identify relevant risks and still fail to govern the system that actually produces them.

We examine this problem through a role-based stress test of the AI RMF. The AI RMF is the object of the stress test, but the broader question applies to governance frameworks more generally: when does framework activity become value-adding governance, and when does it merely generate activity that looks like governance? The test varies organizational role, AI deployment, and governance hard case in a consumer-lending setting. The question is not whether the RMF is useful in general. The question is more specific: under controlled conditions, can RMF language be translated into role-positioned governance reasoning across organizational levels and AI system types?

The study compares two deployments: a bounded Machine Learning (ML) underwriting-risk model and a Large Language Model (LLM) underwriting copilot. The comparison matters because the object of governance may change. In the ML case, the governed object can be treated largely as a model with scores, thresholds, validation, monitoring, and adverse-action reasons. In the LLM copilot case, risk may arise through workflow, reliance, documentation, supervision, and human judgment.

The paper therefore makes an explicit system-boundary move. We treat the object of governance as the AI system-in-use, not only the model or tool artifact. An AI system-in-use includes the model or tool as embedded in work practices, documentation routines, authority structures, monitoring, escalation, and correction. This move foregrounds governability: the degree to which the system-in-use can be known, interpreted, acted on, and corrected from the governance boundary the organization is using. A framework may create value by helping an organization govern. It may also create value by revealing where governability is bounded by current system boundaries, evidence paths, authority structures, or correction capacities.

We analyze this problem through two diagnostic constructs. Structural Fit asks whether the framework is pointed at the right governed object. It asks whether the RMF can attach to the system boundary and operational meaning that matter in use. Governance Translatability asks whether RMF-guided work can move through roles, levels, and authority pathways. It asks whether local framework activity becomes useful governance work across the organization. These diagnostics separate local implementation from governance value. They also separate governance value from downstream risk reduction.

The study uses LLM-based role simulation as a stress test of framework translation into governance over an AI system-in-use under difficult conditions. We vary four organizational roles, two AI deployments, and three governance hard cases. The design produces 24 cells and 120 scored responses. Each response is scored for role understanding, Structural Fit, Governance Translatability, governance value, and plausible risk reduction. The method is intended to expose where framework translation holds, where it breaks, and where framework activity becomes governance-looking work without becoming governance value.

The results show that local translation was not the main problem. Simulated actors could generally use RMF language from their assigned roles. The breaks appeared downstream. Role was associated with governance value because authority affected whether RMF-guided work could reach decisions and return as correction. Deployment was associated with Structural Fit because the RMF fit the bounded ML model more cleanly than the workflow-embedded LLM copilot. Risk reduction was harder to achieve than governance value. It required both strong Governance Translatability and full Structural Fit, plus a concrete mechanism linking governance action to reduced likelihood or impact.

The paper makes two related contributions. For AI governance, it offers a diagnostic account of why framework adoption can fail even when local implementation appears successful. Adoption is not only awareness, training, or use of framework language. It requires framework-guided activity to become authority-connected governance over the AI system-in-use that actually produces risk.

The broader implication is for framework-based AI governance. Frameworks create value only when framework-guided activity can travel through roles, levels, responsibility structures, and authority pathways. They may also create value by making limits of governability visible: what cannot yet be known, what cannot yet be corrected, where the system boundary may need to shift, and where residual risk must be escalated or accepted under current arrangements.

\section{Study Design and Analysis}
\label{sec:study-design-analysis}

This study uses LLM-based role simulation to stress test translation of the NIST AI RMF across organizational roles, AI deployments, and governance hard cases. The purpose is not to predict how real organizational actors would behave. The purpose is to test whether the RMF supports coherent role-positioned governance reasoning under controlled prompt conditions.

\subsection{Experimental design}
\label{subsec:experimental-design}

The study used a balanced factorial design. The design varied four organizational roles, two AI deployments, and three hard cases. This produced 24 experimental cells. We generated five independent replications for each cell, for a total of 120 scored responses.

\begin{table}[htbp]
\centering
\caption{Experimental design}
\label{tab:experimental-design}
\begin{tabular}{lll}
\hline
Factor & Levels & Description \\
\hline
Role & EE, LE, MM, SO & Organizational position \\
Deployment & ML, LLM & AI system type \\
Hard case & A, B, C & Governance stressor \\
\hline
\end{tabular}
\end{table}

The four roles were Enterprise Executive (EE), Line Executive (LE), Middle Manager (MM), and System Owner (SO). The two deployments were a traditional ML underwriting-risk model and an LLM underwriting copilot. The three hard cases were: A, a borderline adverse lending decision outcome with weak justification; B, apparent disparity in lending patterns for a protected class with unclear source and accountability; and C, workflow drift and overreliance.

\subsection{Role-simulation protocol}
\label{subsec:role-simulation-protocol}

Each response was generated from a prompt that assigned one role, one deployment, and one hard case. The model was instructed to respond from the assigned role. It was given the NIST AI RMF as source material and asked to apply the framework to the assigned scenario.

All role-simulation prompts used the same template. Each prompt asked the role actor to address the problem, identify RMF-guided mechanisms, describe the authority path, connect the response to risk, and identify limits. The assigned role, deployment, and hard case were varied by cell; the remaining prompt structure was held constant. We generated five responses per cell using GPT-5.5 with default settings. The generated responses were saved before scoring. The scoring step was conducted separately and did not feed back into generation.

\subsection{Constructs}
\label{subsec:constructs}

The scoring framework rests on a distinction between governance activity, governance value, and risk reduction. Governance activity includes policies, procedures, controls, dashboards, reviews, documentation, committees, and audit evidence. These activities may support governance, but they are not governance by themselves. We use governance to mean the ongoing, rule-mediated coordination of interdependent actors under conditions of limited knowledge to sustain oversight, monitoring, escalation, correction, and mutual legitimacy over time. Governance includes policies, processes, controls, and artifacts, but is not reducible to them.\footnote{This usage draws on institutional and deliberative accounts of governance as rule-mediated coordination, collective action, organizational decision-making, bounded adjustment, and legitimacy. See \cite{north1990institutions}, \cite{ostrom1990governing}, \cite{march1989rediscovering}, \cite{lindblom1959science}, \cite{habermas1990discourse}, and \cite{elster1998deliberative}.} Governance value appears when governance activity improves the organization's ability to see and interpret the system-in-use, assign responsibility, make decisions, escalate signals, authorize correction, defend decisions, preserve value, or prevent loss.

The study also uses governability as the system condition at stake. We use governability to mean the degree to which an AI system-in-use can be known, interpreted, acted on, and corrected from the governance boundary the organization is using. Governability is not scored directly. It explains why the scoring framework asks a sequence of narrower questions: what system is being governed, whether that system has a stable operational meaning, whether the framework fits that system, whether framework-guided work can move through the organization, whether that work creates governance value, and whether it plausibly reduces risk.

Governability is boundary-relative. We therefore make an explicit system-boundary choice. We do not treat the AI model or tool artifact as the only object of governance. We treat the object of governance as the AI system-in-use. This means treating the model or tool as embedded in workflow, human judgment, documentation routines, supervision, incentives, escalation channels, and oversight arrangements. The question is not only whether the RMF can govern a model. The question is whether RMF-guided activity can attach to the system that actually produces organizational and consumer risk.

This boundary choice matters because the same formal AI system can take on different operational meanings in use. Operational meaning refers to the organization's working interpretation of what the AI system is for in practice, what role it plays in workflow, how its outputs should matter, when and how reliance is acceptable, what counts as acceptable performance or failure, and what kinds of correction are appropriate in operation.

Reliance is part of operational meaning in the LLM copilot case. The relevant question is not only whether the tool performs well, but how underwriters interpret, depend on, and act on its outputs. Lee and See's account of trust in automation treats appropriate reliance as central to automation use: automation can help when reliance is warranted, but can create risk when users over-rely, under-rely, or rely under the wrong conditions \citep{lee2004trust}.

The construct definitions below operationalize this premise. To govern an AI system-in-use, an organization must be able to understand the problem from a role, represent what the system is for, describe how it functions in work, evaluate its behavior, identify who has authority over it, and specify how correction occurs. The scoring framework separates one role-positioning construct, two diagnostic constructs, and two downstream judgments. Role Understanding checks whether the response is situated in the assigned role. Structural Fit asks whether the framework fits the system being governed. Governance Translatability asks whether RMF-guided work can move through the organization as governance. Step 1 is a judgment that asks whether that work creates governance value. Step 2 is a judgment that asks whether that value also plausibly reduces risk.

Role Understanding is the first construct. It asks whether the response understands the assigned problem from the assigned role. This is a role-positioning check. It asks whether the actor identifies the relevant problem, explains why it matters from that role, recognizes the role's information position and authority limits, and avoids speaking as a generic AI ethics or governance expert. Role Understanding is not Governance Translatability. It does not ask whether RMF-guided work moves across levels or connects to authority.

Structural Fit asks whether the RMF's implied system model fits the AI system-in-use. It is the governed-object construct. The premise is that system architecture is not only technical structure, but a mapping among system purpose, function, form, and observed behavior. When that mapping provides sufficient constraint, actors can more readily identify what the system is for, what outputs mean, where risk appears, and what correction should target.

This matters because the governed object may not be stabilized by the technical architecture alone. In workflow-embedded AI systems, relevant disturbances may arise through reliance, incentives, documentation routines, institutional interpretation, or distributed authority. These disturbances can destabilize the governed object's operational meaning. In those cases, governance becomes part of the broader organizational constraint regime through which the organization surfaces signals, reconciles interpretations, allocates authority, and coordinates correction.

Structural Fit captures this representational problem. The question is whether the AI system-in-use can be represented across the levels needed for governance: what the system is for, how it functions in work, what outputs mean, where risk appears, who has authority, and what correction can target. Rasmussen's abstraction hierarchy supports this view because complex systems must be represented across levels of abstraction if actors are to understand what is being controlled and what information is decision-relevant \citep{rasmussen1985role}. Zachman similarly supports the need for multiple system representations for different organizational purposes and audiences \citep{zachman1987framework}. Structural Fit therefore asks whether the RMF can attach to the system boundary and operational meaning that matter in use.

Structural Fit has three components: Meaning Determinacy, Closure, and Boundary Containment. Meaning Determinacy asks whether the system's purpose, operational role, acceptable use, and performance criteria are stable enough across actors and contexts to govern. Closure asks whether actual use preserves a stable relationship among intended purpose, design, observed behavior, and evaluation criteria, so the system can still be interpreted and governed in use. Boundary Containment asks whether the sources of risk, harms, evidence, corrective mechanisms, and governing authority remain inside the governance boundary being used.

Governance Translatability asks whether RMF-guided work can move through roles, levels, and authority pathways. Rasmussen's risk-management framework supports this cross-level view of governance: risk management in sociotechnical systems depends on coordination across organizational abstraction levels rather than isolated local action \citep{rasmussen1997risk}. Leveson's systems-theoretic account similarly treats safety and risk as problems of control, feedback, authority, and correction in sociotechnical systems, not only as failures of individual components \citep{leveson2004new}. Governance Translatability therefore asks whether local RMF-guided work becomes usable across levels and connected to authority-backed action.

Governance Translatability has four components: Role Translation, Responsibility Mapping, Cross-Level Governance Value, and Authority Connection. Role Translation asks whether the actor can turn RMF language into governance work available from their role. Responsibility Mapping asks whether the response identifies who owns what, who knows what, who decides, and who acts. Cross-Level Governance Value asks whether role-level work becomes receiver-valued meaning for another organizational level, whether upward to decision-makers or downward to implementation actors. This is not simply a demand for more detail. For upward translation, Cross-Level Governance Value often requires converting verbatim local operational evidence into higher-level gist representations that preserve the decision-relevant meaning of the risk for executives \citep{ReynaBrainerd1995FuzzyTrace,BroniatowskiReyna2018FormalModel}. In practice, those representations are signals tied to risk appetite, resources, strategic priorities, legal or regulatory exposure, customer treatment, operational viability, or correction authority. Authority Connection asks whether the signal reaches someone who can change something and whether the resulting decision, constraint, resource, or correction returns to the actors who must implement it. In plain language, the signal has to answer two questions for the receiving level: so what, and now what?

Step 1 is the governance-value judgment. It asks whether RMF-guided activity becomes governance that matters across organizational levels. This includes improved visibility, sensemaking, responsibility, escalation, decision quality, correction capacity, defensibility, loss prevention, or value preservation. Weick's work on sensemaking supports this part of the scoring logic: under uncertainty, organizations need shared meaning to coordinate and correct action \citep{weick1987organizational, weick1993collapse}. Step 1 is therefore not a count of artifacts. It asks whether RMF-guided activity improves the organization's ability to govern.

Step 2 is the risk-reduction judgment. It asks whether value-adding governance plausibly reduces the likelihood or impact of a relevant organizational or consumer risk pathway associated with the AI system-in-use. Step 2 is downstream from Step 1. Governance value may improve visibility, escalation, responsibility, decision quality, correction, or defensibility without yet showing that likelihood or impact changed. Step 2 requires that further mechanism. It asks how governance action changes model use, human reliance, workflow, documentation, escalation, adverse-action reasoning, fair-lending monitoring, correction, or another relevant pathway.

\subsection{Scoring procedure}
\label{subsec:scoring-procedure}

Scoring was performed in a separate step after response generation. Each response was scored independently by an LLM scorer using the same pre-specified rubric. The scorer evaluated the response text against the assigned role, deployment, hard case, and construct definitions. The scoring instructions required response-specific evidence for full credit. They also directed the scorer not to reward polish, length, confidence, generic RMF fluency, or governance artifacts by themselves. The scorer was instructed to score each response independently rather than from expected role, deployment, hard-case, or cell-level patterns.

The scoring used categorical and ordinal judgments. Component scores used a 0--2 scale. A score of 0 indicated weak or absent evidence. A score of 1 indicated partial evidence. A score of 2 indicated strong evidence. Role Understanding, Role Translation, Responsibility Mapping, Cross-Level Governance Value, Authority Connection, Meaning Determinacy, Closure, and Boundary Containment were scored on this scale.

The scoring procedure kept governance value and risk reduction separate. Governance activity counted as governance value only when it improved organizational visibility, sensemaking, responsibility, escalation, decision quality, correction, defensibility, loss prevention, value preservation, or organizational control. Risk reduction required a further mechanism linking governance action to reduced likelihood or impact.

Governance Translatability was operationalized as a chain. Strong local RMF translation could not compensate for unclear responsibility, weak cross-level value, or a missing authority connection. For that reason, the operational Governance Translatability score was the minimum score across Role Translation, Responsibility Mapping, Cross-Level Governance Value, and Authority Connection.

Structural Fit was scored as an equally weighted composite of Meaning Determinacy, Closure, and Boundary Containment. For the primary binary tables, Full Structural Fit meant a Structural Fit score of 2.00. Responses below 2.00 were coded as not full.

Step 1 and Step 2 were assigned as summary judgments. Each was coded as Yes, Partial, or No. Step 1 asked whether RMF-guided activity produced governance value. Step 2 asked whether that governance value plausibly reduced likelihood or impact through a concrete risk-reduction mechanism.

Visible Value was coded as a separate adjudication. It distinguished responses that showed visible governance value from responses that were mixed, unclear, or mostly process, controls, and artifacts.

For the primary contingency tables, the main outcomes were converted to binary categories. Scores of 2 were coded as Strong, and scores below 2 were coded as Not strong. Step 1 and Step 2 Yes judgments were coded as the positive category. Partial and No judgments were combined as the reference category. Visible Value was coded as the positive category, while Mixed/Unclear and Mostly process/controls/artifacts were combined as the reference category.

\subsection{Outcome coding}
\label{subsec:outcome-coding}

The main outcomes were categorical or ordinal. For the primary contingency tables, we used binary versions of the main outcomes. This kept the analysis focused on whether a response reached the full bar for each construct.

\begin{table}[htbp]
\centering
\caption{Primary binary outcome coding}
\label{tab:binary-coding}
\begin{tabular}{lll}
\hline
Construct & Positive category & Reference category \\
\hline
Cross-Level Value & Strong & Not strong \\
Authority Connection & Strong & Not strong \\
Governance Translatability & Strong & Not strong \\
Step 1 & Yes & Partial/No \\
Step 2 & Yes & Partial/No \\
Visible Value & Visible Value & Not Visible Value \\
Closure & Strong & Not strong \\
Boundary Containment & Strong & Not strong \\
Structural Fit & Full Structural Fit & Not full \\
\hline
\end{tabular}
\end{table}

Partial and No judgments were combined as the reference category in the binary tables. This coding does not treat Partial and No as conceptually identical. It only distinguishes responses that reached the full bar from responses that did not.

\subsection{Statistical analysis}
\label{subsec:statistical-analysis}

The analysis used contingency tables. For each primary table, we report observed counts and proportions, Pearson's chi-square test, the associated p-value, and Cramer's $V$. Cramer's $V$ is reported as the association-strength measure. Where expected counts were small or cells contained zeros, we used exact-test checks. Fisher's exact test was used for 2 by 2 tables. Fixed-margin Monte Carlo exact checks were used for larger sparse tables.

The main tests followed the study claims. Role was tested against governance-value outcomes. Deployment was tested against Structural Fit outcomes. Deployment was also tested against Step 1 and Step 2. Hard-case-by-deployment tables were used as diagnostic follow-up analyses.

\subsection{Scope of inference}
\label{subsec:scope-inference}

The analysis supports claims about the scored outputs from this LLM role-simulation protocol. The replications within each cell were treated as independent draws from the model response distribution conditional on the assigned prompt condition. Across cells, responses were not treated as identically distributed because the role, deployment, and hard case changed.

\section{Results}
\label{sec:results}

The results support the main story of the paper. Local translation was easy in the simulation. Getting to governance value was hard. Risk reduction was harder still. The RMF did not break at local use. The simulated actors understood the prompt, stayed in role, and translated RMF language into local activity. The problem appeared when that local activity had to become governance. It had to move across organizational levels and connect to authority for decisions. Those decisions had to return to the lower levels of the organization for implementation.

\subsection{Local translation was not the problem}
\label{subsec:local-translation}

The first result is that local translation was not the bottleneck. Role Understanding, Role Translation, and Responsibility Mapping were at ceiling: all 120 responses received full credit on each of these constructs. The simulated actors could identify the problem. They could speak about the problem imposed by the hard case from their assigned role. They could describe RMF-guided activity that made sense locally. They also understood who in the organization was empowered to make decisions in response to the issues the role-actor identified. It should be noted that this is probably an overly optimistic view of role-actors' situational awareness compared to their counterparts in real organizations. The LLM-simulated role-actors are unusually aware of the organizational landscape, its players, and their roles / job duties. In that sense, this is a conservative test of our theoretical arguments. Even with unusually high situational awareness, local RMF activity still did not always become governance value or risk reduction.

This matters because it rules out the simplest explanation for weak adoption. The RMF did not fail because the actors could not understand it at all or make sense of who in their leadership chain could make decisions around the identified issue. The framework was locally usable under the conditions of the study and role-actors understood who had decision rights and needed information. The harder question was whether local use and organizational knowledge could be translated into governance. For that to happen, RMF-guided work had to become useful to other roles, connect to authority, and support a decision, escalation, correction, or risk acceptance.

\subsection{Role drove governance value}
\label{subsec:role-governance-value}

Role was strongly associated with governance-value outcomes. Table~\ref{tab:role-governance} shows the pattern. Enterprise Executive (EE) responses were at ceiling across the governance-value outcomes. Line Executive (LE) responses were lower, but still much stronger than Middle Manager (MM) and System Owner (SO) responses. The main split was not between actors who understood the RMF and actors who did not. It was between roles that had the means to be able to connect RMF-guided work to authority and roles that could not.

\begin{table}[htbp]
\centering
\caption{Role by governance-value outcomes}
\label{tab:role-governance}
\begin{tabular}{lccccccc}
\hline
Outcome & EE & LE & MM & SO & $\chi^2$ & $p$ & Cramer's $V$ \\
\hline
Cross-Level Value strong & 30/30 & 23/30 & 6/30 & 4/30 & 65.33 & $< .000001$ & .738 \\
Authority Connection strong & 30/30 & 19/30 & 6/30 & 3/30 & 62.07 & $< .000001$ & .719 \\
Governance Translatability strong & 30/30 & 18/30 & 6/30 & 3/30 & 61.05 & $< .000001$ & .713 \\
Step 1 Yes & 30/30 & 18/30 & 6/30 & 3/30 & 61.05 & $< .000001$ & .713 \\
Visible Value & 30/30 & 18/30 & 6/30 & 3/30 & 61.05 & $< .000001$ & .713 \\
\hline
\end{tabular}
\end{table}

This should not be read as evidence that executives reasoned better. It is better read as evidence about the structure of authority. In most organizations, authority is concentrated at higher levels and delegated downward. Enterprise and Line Executives could connect a risk signal to risk appetite, resources, stop-use decisions, remediation, or residual-risk acceptance. Middle Managers and System Owners often produced useful local machinery. But local machinery is not enough. It becomes governance value only when it reaches authority and returns as correction.

The reverse problem also matters. Implementation responsibility is often pushed downward to Middle Managers and individual contributors, represented by the System Owner in this study. These actors are closer to actual use, so they may see workflow drift, reliance problems, and documentation gaps first. But they may not have the authority to change use, pause a system, accept residual risk, or allocate resources for correction. A breakdown in Governance Translatability therefore blocks both directions. Signals do not travel up in a form senior actors can use, and decisions do not travel down in a form local actors can implement.

One final thing to note, in this scoring round, Governance Translatability, Step 1, and Visible Value moved together. We should not treat them as three independent discoveries. They are related diagnostics of the same governance-value pathway. The narrower interpretation is stronger. Governance Translatability was the binding condition for Step 1—creating governance value. Local RMF work had to become receiver-valued and authority-connected before it counted as governance value.

\subsection{Deployment drove Structural Fit}
\label{subsec:deployment-structural-fit}

System type also mattered. The RMF fit the bounded ML model cleanly. It fit the LLM copilot less well. Table~\ref{tab:deployment-sf} shows the difference. The ML underwriting model was at ceiling on Closure, Boundary Containment, and Full Structural Fit. The LLM copilot was not.

\begin{table}[htbp]
\centering
\caption{Deployment by Structural Fit outcomes}
\label{tab:deployment-sf}
\begin{tabular}{lccccc}
\hline
Outcome & ML & LLM & $\chi^2$ & $p$ & Cramer's $V$ \\
\hline
Closure strong & 60/60 & 37/60 & 28.45 & $< .000001$ & .487 \\
Boundary Containment strong & 60/60 & 34/60 & 33.19 & $< .000001$ & .526 \\
Full Structural Fit & 60/60 & 33/60 & 34.84 & $< .000001$ & .539 \\
\hline
\end{tabular}
\end{table}

The ML case gave the RMF a clearer object to govern. The risk-scoring model had familiar points of control: scores, thresholds, validation, monitoring, overrides, adverse-action reasons, drift, and fair-lending testing. These points of control made it easier for RMF-guided activity to connect to evidence, decisions, and correction.

The LLM copilot presented a different problem. The governed object was not just the model. It was the model as used inside underwriting work. The relevant risks came through workflow, reliance, narrative framing, documentation practice, supervision, and human judgment. The RMF could still generate activity in this setting, but it was harder to know where governance should attach and what correction should change.

The decomposed Structural Fit scores point to the same issue. Meaning Determinacy was at ceiling: all 120 responses received full credit on this component. This likely reflects the prompt design. The prompt made the system purpose, role, and hard case clear. Actors knew what the system was supposed to do. The real variation came from Closure and Boundary Containment. These are harder tests. They ask whether actual use stayed connected to intended use, and whether the organization could still see, interpret, and correct the risk from inside its governance boundary. The LLM case was weaker on these tests because the risk moved through workflow, reliance, documentation, and human judgment. The problem was not understanding the stated purpose of the system. The problem was governing what the system became in use.

\subsection{Governance value was easier than risk reduction}
\label{subsec:step1-step2}

Step 1—governance value—and Step 2—risk reduction—separated clearly. Deployment did not explain whether RMF-guided activity became governance value. A "Yes" for Step 1, indicating governance value, appeared at similar rates in the ML and LLM deployments. But deployment did help explain whether governance value became plausible risk reduction. A "Yes" for Step 2, indicating that risk was judged to be reduced, appeared more often in the ML deployment. Table~\ref{tab:deployment-step} shows this pattern.

\begin{table}[htbp]
\centering
\caption{Deployment by Step 1 and Step 2}
\label{tab:deployment-step}
\begin{tabular}{lccccc}
\hline
Outcome & ML & LLM & $\chi^2$ & $p$ & Cramer's $V$ \\
\hline
Step 1 Yes & 27/60 & 30/60 & 0.30 & .583 & .050 \\
Step 2 Yes & 17/60 & 5/60 & 8.02 & .0046 & .258 \\
\hline
\end{tabular}
\vspace{0.5em}
\begin{flushleft}
\footnotesize
Note: For Step 1, Fisher's exact $p=.715$. For Step 2, Fisher's exact $p=.0084$. Partial and No judgments are combined in the reference category.
\end{flushleft}
\end{table}

This matters because it shows where system type enters the results. Governance value was not simply an ML effect. The LLM condition could also produce governance value when local RMF work connected to authority. Risk reduction was different. Full credit for Step 2 depended on more than authority-connected governance. It also depended on whether the RMF fit the system being governed. This points back to Structural Fit. The ML case gave the RMF clearer points of control (e.g., model scores). The LLM case made the governed object harder to stabilize, bound, and correct.

Table~\ref{tab:step1-step2} shows the direct relationship between Step 1 and Step 2. All Step 2 "Yes" judgments occurred among responses that received full credit for Step 1. But most Step 1 "Yes" judgments did not translate into the response getting full credit for Step 2. This supports the central distinction. Governance value is broader than risk reduction. Governance can improve visibility, escalation, responsibility, decision quality, and correction capacity without showing that likelihood or impact changed.

\begin{table}[htbp]
\centering
\caption{Step 1 by Step 2}
\label{tab:step1-step2}
\begin{tabular}{lccc}
\hline
Step 1 status & Step 2 Partial/No & Step 2 Yes & Total \\
\hline
Step 1 Partial/No & 63 & 0 & 63 \\
Step 1 Yes & 35 & 22 & 57 \\
\hline
Total & 98 & 22 & 120 \\
\hline
\end{tabular}
\vspace{0.5em}
\begin{flushleft}
\footnotesize
Note: Pearson $\chi^2(1)=29.77$, $p<.000001$; Fisher's exact $p<.000001$; Cramer's $V=.498$. Because Step 1 and Step 2 are paired judgments on the same responses, an exact McNemar test was also used; $p=5.8\times10^{-11}$. Partial and No judgments are combined in the reference category.
\end{flushleft}
\end{table}

The next question is what separated cases that produced governance value (Step 1) from those that also achieved risk reduction (Step 2). The first answer is Governance Translatability. Table~\ref{tab:gt-step2} shows the relationship between Governance Translatability and risk reduction (Step 2). No response received full credit for risk reduction unless Governance Translatability was strong. But strong Governance Translatability did not guarantee risk reduction. Of the 57 responses with strong Governance Translatability, only 22 received a "Yes" for risk reduction (Step 2).

\begin{table}[htbp]
\centering
\caption{Governance Translatability by Step 2}
\label{tab:gt-step2}
\begin{tabular}{lccc}
\hline
Governance Translatability & Step 2 Yes & Step 2 Partial/No & Step 2 Yes among group \\
\hline
Strong & 22 & 35 & 22/57 \\
Not strong & 0 & 63 & 0/63 \\
\hline
Total & 22 & 98 & 22/120 \\
\hline
\end{tabular}
\vspace{0.5em}
\begin{flushleft}
\footnotesize
Note: Pearson $\chi^2(1)=29.77$, $p<.000001$; Fisher's exact $p<.000001$; Cramer's $V=.498$. Partial and No judgments are combined in the reference category.
\end{flushleft}
\end{table}

This table puts the first stake in the ground. Risk reduction did not appear without Governance Translatability. Local RMF work had to become cross-level and authority-connected before it could plausibly reduce risk. But Governance Translatability was not enough. Most responses with strong Governance Translatability still did not receive full credit for risk reduction. The next question is what separated the strong Governance Translatability cases that reached the risk reduction bar for Step 2 from those that did not.

Table~\ref{tab:conditional-step2} conditions on a Yes for Step 1 (full credit for governance value) and compares risk reduction (Step 2) outcomes by deployment and Structural Fit. This is the more direct test of whether governance value converted into plausible risk reduction.

\begin{table}[htbp]
\centering
\caption{Step 2 among Step 1 Yes responses by deployment and Structural Fit}
\label{tab:conditional-step2}
\begin{tabular}{lccc}
\hline
Deployment / Structural Fit group & Step 2 Yes & Step 2 Partial/No & Step 2 Yes among Step 1 Yes \\
\hline
ML + full Structural Fit & 17 & 10 & 17/27 \\
LLM + full Structural Fit & 5 & 6 & 5/11 \\
LLM + not-full Structural Fit & 0 & 19 & 0/19 \\
\hline
\end{tabular}
\vspace{0.5em}
\begin{flushleft}
\footnotesize
Note: Pearson $\chi^2(2)=18.93$, $p=.000078$; fixed-margin Monte Carlo exact $p<.0001$; Cramer's $V=.576$. This table conditions on responses that received Step 1 Yes. Partial and No judgments are combined in the reference category.
\end{flushleft}
\end{table}

The pattern is clear. When Structural Fit was not full, governance value did not translate into risk reduction. LLM responses without full Structural Fit received full credit for governance value in 19 cases, but none received full credit for risk reduction. Full Structural Fit helped, but it was not enough alone. A "Yes" judgment for Step 2 still depended on whether the response linked governance action to reduced likelihood or impact.

Together, the two tables state the main result about risk reduction. Governance Translatability was necessary for a response to receive full credit for Step 2 because risk reduction did not appear unless RMF-guided work became cross-level and authority-connected. Structural Fit was also necessary because risk reduction (Step 2) did not appear when Structural Fit was not full. Neither condition was sufficient. Risk reduction still required a concrete mechanism linking governance action to reduced likelihood or impact.

\subsection{Hard cases explain where the breaks occurred}
\label{subsec:hard-cases}

The hard-case patterns help explain the main results. They should be treated as diagnostic, not as the main statistical claim. Hard Case B was the strongest governance trigger. This makes sense. Apparent disparity activates familiar pathways. It points to fair lending, legal exposure, compliance, model risk, consumer harm, and business legitimacy. The threats contained in those signals are clear across levels throughout the organization.

Hard Case C was different. Workflow drift was visible, but it was hard to convert into risk reduction. This was clearest in the LLM copilot condition. LLM\_C had relatively high Full Structural Fit, but zero Step 2 "Yes" judgments. That is an important distinction. Seeing workflow drift is not the same as correcting it. The system-in-use problem may be visible before it is authority-connected and correctable.

\begin{table}[htbp]
\centering
\caption{Hard case and deployment diagnostic summary}
\label{tab:hardcase-deployment}
\begin{tabular}{lcccc}
\hline
Condition & GT strong & Step 1 Yes & Step 2 Yes & Full Structural Fit \\
\hline
ML\_A & 8/20 & 8/20 & 6/20 & 20/20 \\
ML\_B & 13/20 & 13/20 & 8/20 & 20/20 \\
ML\_C & 6/20 & 6/20 & 3/20 & 20/20 \\
LLM\_A & 8/20 & 8/20 & 1/20 & 9/20 \\
LLM\_B & 15/20 & 15/20 & 4/20 & 8/20 \\
LLM\_C & 7/20 & 7/20 & 0/20 & 16/20 \\
\hline
\end{tabular}
\end{table}

This table also shows why Structural Fit is not enough. LLM\_C made the system-in-use problem easier to see. But it still did not produce Step 2 risk reduction. Visibility alone did not become correction. This finding connects the hard-case analysis back to the larger governance problem. Local actors may see the problem before the organization has a clear path to act on it. The organization may still fail to act if authority connections between organizational levels are not strong enough.

\section{Discussion}

The results suggest that the RMF adoption problem identified by \cite{kolo2025framework} reflects two interacting abstraction mismatches. The first is organizational. RMF-guided work must move across roles that differ in authority, information, incentives, bandwidth, and definitions of value. The second is system-based. RMF-guided work must fit AI deployments that differ in how much the system architecture stabilizes operational meaning. These mismatches help explain why the RMF can be locally usable while still failing to produce governance value or risk reduction.

The organizational mismatch is a conversion problem. Authority tends to move top-down, while actionable information often moves bottom-up. Enterprise and Line Executives hold much of the authority to set priorities, allocate resources, constrain use, accept residual risk, or pause deployment. Middle Managers and System Owners often build the machinery through which governance is implemented. Implementation therefore concentrates where authority is thinner, while authority concentrates where operational detail is thinner.

This pattern explains both the strong role effects and the central importance of authority connections. Senior-authority roles were more likely to close the governance-value pathway because they could connect RMF-guided work to risk appetite, resources, decisions, and correction, provided the signal reached them. Lower-authority roles often saw important local issues, but they could not always make those issues consequential. This is the visibility-authority problem. Local actors may see the risk first, while senior actors are better positioned to act on it. For governance to work, information and decisions must traverse the organizational abstraction hierarchy. Signals must move upward in a form decision-makers can use, and decisions must move downward as constraints, resources, or correction. Those authority connections come from the organization's existing governance capability maturity, or they must be deliberately built.

This matters for adoption because local implementation can look successful while governance still fails to create value or reduce risk exposure. Middle managers, system owners, and other individual contributors can produce controls, monitoring routines, documentation, and escalation language. Yet those activities become governance value only when they reach a decision point in a form decision makers can integrate, and return as correction. Adoption is therefore partly a question of whether staff can use the RMF, but mostly a question of whether the organization has enough governance maturity to make RMF-guided work matter. This may help explain the awareness-adoption gap: organizations may know the RMF, but still lack the authority connections needed to turn framework activity into value.

The system mismatch concerns the object being governed. The RMF fit the bounded ML model cleanly because the model contribution was comparatively stable and legible. The underwriting-risk model produced scores, thresholds, validation results, drift signals, override patterns, adverse-action reasons, and fair-lending indicators. These are familiar points of control. They can be connected to existing model-risk, compliance, escalation, and correction routines. In that setting, the central governance question is whether the model is behaving within an approved control envelope.

The LLM copilot presented a different governance problem. Labeling the system as advisory does not settle what role it actually plays in underwriting work. In practice, it may shape what underwriters notice, how they frame ambiguity, which facts become salient, how rationales are assembled, and how nominally human judgment is exercised. The model may produce an inaccurate output, but that is not the only issue. The larger issue is that the system's practical role in the workflow may be reinterpreted through use—its operational meaning shifts. Whether the model is correct remains relevant, but it is not enough. The more important governance question is what role the system is actually playing in organizational judgment, whether that role is acceptable, and whether the organization can constrain or redesign the workflow when actual use diverges from intended use.

This is why deployment mattered for Structural Fit. Workflow-embedded LLM systems strain Closure and Boundary Containment because risk moves through reliance, documentation, supervision, and human judgment. Monitoring and enforcement remain necessary, but they are not sufficient. Governance also has to support sensemaking about actual use. It has to make weak signals visible, interpret reliance patterns, review workflow drift, and connect those signals to authority-backed correction. For the AI RMF, this means that guidance built around lifecycle controls and model-centered evidence may fit bounded ML systems more readily than LLM systems whose risks emerge through use.

The two mismatches compound. The less stable and observable the system's operational role becomes, the more work the organization must do to operate at the limits of the system's governability. Local signals have to be found, interpreted, translated upward, connected to authority, and returned as correction. That work is the cost imposed by the abstraction mismatch. It appears differently at different levels. Executives have to invest more attention and cognitive effort in sensemaking and harder decisions. Line executives have to manage workflow redesign, productivity tradeoffs, customer-treatment accountability, and operating constraints. Middle managers, system owners, and individual contributors have to produce local evidence, maintain monitoring routines, and turn operational observations into signals executive customers can use.

In the ML case, signals are already recognizable within the existing governance and compliance setting. In the LLM case, signals are considerably harder to observe, measure, and translate into executive-legible decisions with the tools available. If the organization treats the LLM copilot as a bounded model when the relevant risk is emerging through workflow, governance attaches to the wrong boundary. More RMF activity may then produce more artifacts without improving control or clarifying the limits of governability: what the organization can know and correct from that boundary. The results suggest a dependency that future work should test directly: the less structurally legible the governed system is, the more governance capability and boundary work are needed to make RMF activity value-adding.

This clarifies what RMF adoption can and cannot be expected to do. Governance value is not the same thing as risk reduction. RMF-guided work may improve visibility, sensemaking, decision quality, correction capacity, loss prevention, or value preservation before it can show a concrete reduction in likelihood or impact. Those gains matter. They are part of what makes a governance framework useful. But they should not be confused with risk reduction. A stronger risk-reduction claim requires a further showing: how governance action changes what people do, what the system does, or what the organization can detect and correct.

This distinction helps explain why serious RMF adoption may be unattractive inside organizations. The easiest parts of adoption are often the most visible: meetings, dashboards, reviews, policies, documentation, and evidence packages. The harder parts are the parts that change organizational behavior: constraining use, reallocating authority, redesigning workflow, slowing deployment, escalating residual risk, or accepting that current governance boundaries are inadequate. Organizations can therefore appear to adopt the RMF while avoiding the more costly work required to make RMF activity consequential. That is how framework adoption becomes compliance activity: formally present, operationally weak, and easy to abandon.

For NIST, the implication is that RMF implementation guidance should focus less on artifact production and more on governance capability. A useful playbook would connect each RMF activity to the decision it improves, the signal it produces, the actor who interprets that signal, the authority needed to act, and the corrective options available. It should also provide role-specific value propositions. Executives, line leaders, middle managers, and system owners need to see what RMF value looks like from their role and how their role contributes to cross-level governance value.

NIST guidance should also help organizations distinguish system types. Some AI deployments present comparatively bounded governance objects. Outputs, thresholds, validation results, drift metrics, adverse-action reasons, and disparity signals can be connected to existing control and escalation mechanisms. Other deployments, especially workflow-embedded LLM systems, generate risk through reinterpretation of system role, reliance, narrative framing, documentation practices, and distributed corrective authority. The point is not to create a rigid taxonomy. The point is to help organizations recognize when monitoring and enforcement are likely to be enough, and when governance must do more sensemaking work to stabilize the system's operational role.

This study has three main limits. First, it uses LLM-based role simulation rather than observed organizational teams. Second, it studies one domain, consumer lending, and two stylized AI deployments. Third, the scoring is diagnostic and theory-building rather than a validated measurement instrument. These limits point directly to the next stage of the work: testing the protocol with human subjects or organizational teams, extending it to additional domains and AI system types, using multiple raters, and varying organizational governance maturity. The most important next step is to test whether organizations with stronger governance capability are better able to convert RMF activity into value-adding governance rather than compliance artifacts.

Taken together, the findings suggest that framework adoption is partly a problem of awareness, training, and implementation, but first and foremost a governance problem. Frameworks are adopted when they create value inside real governance systems, not when they merely generate framework-shaped activity. This point applies to the AI RMF specifically, but it also generalizes to governance frameworks more broadly. Frameworks have to travel through roles, responsibilities, levels, and authority pathways. For the AI RMF, adoption therefore depends on whether organizations can connect local use to authority, and whether the framework is pointed at the AI system-in-use that actually creates the risk.

\section{Conclusion}

This paper contributes a diagnostic account of AI governance framework adoption. It shows that adoption is not only a matter of awareness, training, or local implementation. Framework activity must become authority-connected governance over the AI system-in-use. The paper captures this through Governance Translatability and Structural Fit, and separates governance value from downstream risk reduction.

The findings speak first to the AI RMF. The RMF can support local role reasoning, but local use is not enough. RMF-guided activity creates governance value only when it moves across organizational levels, connects to authority, and attaches to the system-in-use that actually produces risk. For governance to work, information and decisions must traverse the organizational abstraction hierarchy. Signals must move upward in a form decision-makers can use, and decisions must move downward as constraints, resources, or correction.

The findings also show why workflow-embedded AI systems create a harder governance problem. For bounded ML systems, governance can often attach to familiar points of control. For LLM copilots, the governed system may extend beyond the model. In those cases, the question is not only whether the model is performing correctly. It is what role the system is actually playing in organizational judgment, whether that role is acceptable, and whether the organization can correct it when actual use diverges from intended use.

The findings also speak to governance frameworks more generally. Frameworks have to travel through real organizations. They create value when they improve visibility, sensemaking, responsibility, escalation, decision quality, correction capacity, defensibility, value preservation, or loss prevention. They do not create value merely by generating framework-shaped activity. Serious adoption is therefore costly. It can require new authority connections, workflow redesign, stronger evidence paths, slower deployment, residual-risk escalation, or a shift in the system boundary being governed.

Finally, frameworks can create value by revealing limits of governability. A framework may not eliminate uncertainty, create missing authority, make opaque systems transparent, or produce correction capacity by itself. But it can help an organization locate the ceiling: what cannot yet be known, what cannot yet be corrected, where the system boundary may need to shift, where reliance may need to be bounded, and where residual risk must be escalated or accepted. That diagnostic function is central to making AI governance frameworks useful in practice.

\subsubsection*{Acknowledgements.}
This work was supported by the NIST-NSF Institute for Trustworthy AI in Law and Society (TRAILS), under Award No. 2229885. Any opinions, findings, and conclusions or recommendations expressed in this material are those of the author(s) and do not necessarily reflect the views of the NSF or NIST.

\bibliographystyle{unsrtnat}
\bibliography{references}

\end{document}